\documentclass[traditabstract]{aa}
\usepackage{times}
\usepackage{graphics,graphicx,natbib}
\bibpunct{(}{)}{;}{a}{}{,}

\def\pdot {\dot P}

\def\ltsima{$\; \buildrel < \over \sim \;$}
\def\lsim{\lower.5ex\hbox{\ltsima}}
\def\gtsima{$\; \buildrel > \over \sim \;$}
\def\gsim{\lower.5ex\hbox{\gtsima}}

\def\xmm{{\em XMM--Newton}}

\def\hd {HD\,49798}
\def\rx {RX\,J0648.0--4418}
\def\hr {HD\,49798/RX\,J0648.0--4418}

\begin{document}

\title
{X-ray emission from the luminous O-type subdwarf HD 49798 and its compact companion}
\author{S. Mereghetti\inst{1}, N. La Palombara\inst{1}, A. Tiengo\inst{2,1,4}, N. Sartore\inst{1},
P. Esposito\inst{1}, G.L. Israel\inst{3}, L.Stella\inst{3}}

\institute{INAF, Istituto di Astrofisica Spaziale e Fisica
Cosmica Milano, via E.\ Bassini 15, I-20133 Milano, Italy
\and
IUSS, Istituto Universitario di Studi Superiori, piazza della Vittoria 15, I-27100 Pavia, Italy
\and
INAF, Osservatorio Astronomico di Roma, via Frascati 33, I-00040 Monteporzio Catone, Italy
\and
INFN, Istituto Nazionale di Fisica Nucleare, Sezione di Pavia, via Bassi 6, I-27100 Pavia, Italy
}

\offprints{S. Mereghetti, sandro@iasf-milano.inaf.it}

\date{Received February 9, 2013 / Accepted April 4, 2013}

\authorrunning{S. Mereghetti et al.}

\titlerunning{X-rays from the sdO HD 49798 and its compact companion}

\abstract {The X-ray source \rx\ is the only confirmed binary system in which a compact object, most likely
a massive white dwarf, accretes from a hot subdwarf companion, the bright sdO star \hd .
The X-ray emission from this system is characterized by two periodic modulations  caused by
an eclipse, at the orbital period of 1.55 d, and by the rotation of the compact object with a spin period of 13.2 s.
In 2011 we  obtained   six short \xmm\  observations centered at orbital phase 0.75, in order to study the system during the eclipse,
and spaced at increasingly long time intervals   in order to obtain
an accurate measure of the spin-period evolution through phase-connected timing.
The duration of the eclipse ingress and egress, $\sim500$ s, indicates the presence of an X-ray  emitting region
with dimensions of the order of a few 10$^4$ km,
surrounding the pulsar  and probably due to scattering in the companion's wind.
We derived an upper limit on the  spin-period derivative  $|\pdot|<6\times10^{-15}$ s s$^{-1}$, more than
two orders of magnitude smaller than the previously available value.
Significant X-ray emission is detected also during the 1.2 hours-long eclipse,
with a luminosity  of $\sim$3$\times$10$^{30}$ erg s$^{-1}$.
The eclipse spectrum shows prominent emission lines  of H- and He-like nitrogen,
an overabundant element in \hd .
These findings support the suggestion that the X-ray emission observed during the eclipse originates in
\hd\   and that the processes responsible for X-ray emission in the stellar winds
of   massive O stars are also at work in the much weaker winds of hot subdwarfs.
}

\maketitle

\section{Introduction}

The   X-ray source \rx\ is
one of the very few known binaries composed of
a hot subdwarf and a compact object.  The optical component is the
bright blue star \hd ,   classified as a subdwarf of spectral type O6,  with  effective temperature
$T_{\mathrm{eff}} = 47,500$ K, and surface gravity $\log g = 4.25\pm0.2$ \citep{kud78}.
Early spectroscopic observations  showed a dominance of He and N lines and an underabundance of C and O,  as well as
radial velocity variations  pointing to a binary nature, that was later confirmed through the
discovery of an orbital period of 1.5477 d \citep{tha70}.

The nature of the  companion   of \hd ,  undetectable in the optical band due to the presence of the much brighter sdO star,
could be revealed only in 1996, thanks to  the \textsl{ROSAT} discovery of soft X-rays pulsed with a period of 13.2 s \citep{isr97},
compatible only with either a neutron star or a white dwarf.
While with the \textit{ROSAT} data it was not possible to discriminate between these two possibilities,
more recent observations carried out with \xmm\ showed that the compact object is most likely a massive white dwarf \citep{mer09}.

For a  distance of 650 pc\footnote{The poorly constrained Hipparcos parallax of 1.16$\pm$0.63 mas \citep{per97} is  consistent with this distance.}  \citep{kud78}, the observed X-ray flux  corresponds to a bolometric X-ray luminosity of
$\sim$10$^{32}$ erg s$^{-1}$.
Taking into account the relatively well known properties of the stellar wind of \hd\ \citep{ham10}, it can be
shown that this luminosity is consistent with that expected for Bondi-Hoyle accretion onto a white dwarf,
but too small for an accreting neutron star \citep{mer11}.
The presence of an X-ray eclipse lasting $\sim$1.2 h indicates
that  the system is seen at  high inclination $i$ $\sim82^{\circ}$.
The masses of the two components have been dynamically  measured through  X-ray pulse timing and the spectroscopically
determined optical mass function:
they are  $M_{\mathrm{opt}} =1.50\pm0.05~M_\odot$ for the subdwarf  and $M_{\mathrm{X}} = 1.28\pm0.05~M_\odot$ for its companion \citep{mer09}.
Currently,   \hd\ is well within its Roche-lobe and   mass transfer through wind accretion proceeds at a low rate
of $\sim$$8\times10^{-13}$ $M_\odot$ yr$^{-1}$.
After the formation of a CO core, the subdwarf will expand
and transfer He-rich material through Roche-lobe overflow at a higher rate.
If the compact object is indeed a  massive  white dwarf, it may reach the Chandrasekhar  limit and either explode as a type Ia supernova \citep{wan10}
or form a fast-spinning neutron star through an accretion-induced collapse.
This could be a  way to create a millisecond
pulsar directly, i.e. without a spin-up phase in a low mass X-ray binary.

During the only X-ray eclipse observed in May 2008 some faint X-ray emission was   detected, but it could
not be studied in detail with the short available exposure. We therefore carried out a series of
time constrained  \xmm\  observations
with the main objective to study the X-ray emission during the eclipse of this unique binary.

\begin{table*}[htbp!]
\caption{Log of the 2011  \xmm\ observations of \rx.
\label{table_obs}}
\begin{tabular}{ccccccc}
\hline\hline
Observation &  Date & MJD start  & MJD end & Orbital Phase$^a$ & Duration pn/MOS (ks) \\
\hline
1 & 2011 May 02 &  55683.550  &   55683.764  & 0.685--0.824   & 17.0 / 18.5 \\
2 & 2011 Aug 18 &  55791.876  &   55792.069  & 0.677--0.803   & 15.0 / 16.6 \\
3 & 2011 Aug 20 &  55793.458  &   55793.624  & 0.699--0.808   & 11.8 / 14.3\\
4 & 2011 Aug 25 &  55798.040  &   55798.267  & 0.660--0.806   & 18.0 / 19.6 \\
5 & 2011 Sep 03 &  55807.349  &   55807.542  & 0.675--0.801   & 15.0 / 16.6 \\
6 & 2011 Sep 08 &  55811.997  &   55812.189  & 0.678--0.802   & 15.0 / 16.6 \\
\hline
\end{tabular}
\begin{small}
\\
$^{a}$ Phase 0.75 corresponds to the middle of the eclipse.
\\
\end{small}
\end{table*}

\section{Observations and data reduction}

Six \xmm\ observations of \hr\ were performed between May and September 2011. Each observation lasted about 15 ks and included
orbital phase 0.75, corresponding to the center of the X-ray eclipse
(see Table \ref{table_obs} for details).
Here we report on the results obtained with  the EPIC (0.15--12 keV) and RGS  (0.3--2.5 keV) instruments.
The data were  processed using version 11 of the \xmm\ Standard Analysis Software.
Some observations showed the presence of short time intervals affected by a high flux of soft protons,
which caused a slight  increase in the particle background    at high energies.
We verified that our spectral results are insensitive to the inclusion or exclusion
of such time intervals and therefore decided to use the whole data set.

EPIC consists of two MOS and one pn CCD cameras \citep{tur01,str01}.
During all the observations the three cameras were operated in Full Frame mode
(time resolution of 73 ms and 2.6 s for pn and MOS, respectively)
and with the medium optical blocking filter.
For the timing analysis we used a circular extraction region  with radius
of 30$''$.  The  photon  arrival
times were converted  to the Solar System barycenter reference frame, by using the
coordinates $\rm R.A. = 6^h\,48^m\,04.7^s$, $\rm Decl.=-44^\circ\,18'\,58.4''$ (J2000),
and corrected for the orbital motion of the source with the system parameters given in \citet{mer11}.
In order to avoid possible contamination  from a nearby source (located 70$''$ away), we extracted the counts for the EPIC spectral
analysis from  a  circle of 20$''$ radius, which corresponds to an enclosed energy fraction of $\sim$80\% and $\sim$75\%
for the pn and MOS, respectively.
The spectra were rebinned so as to have at least 30 counts per energy channel.
The background spectra were extracted from source-free regions on the same chip as the target.
The spectral fits were done with version 12.7 of the XSPEC package.

The RGS instrument \citep{denHerder+01long}  has a
significantly smaller effective area than EPIC, but its excellent
spectral resolution makes it particularly sensitive to narrow emission
lines even in relatively faint X-ray sources like \rx . We extracted
the first order spectra and response matrices from both  RGS units
in each observation using
a source extraction region corresponding to 80\% of the point spread function in the cross
dispersion direction. This region is smaller than that typically
adopted in RGS spectral analysis and was chosen in order to minimize
the contamination from the nearby source mentioned above (very likely
an active star), which shows prominent X-ray emission lines. The
source and background spectra and the response matrices of the seven
observations were combined together to obtain a cumulative spectrum for each RGS camera.
These spectra were rebinned in order to have at least 20 counts per energy channel.

 \begin{figure}
 \includegraphics[width=45mm,angle=-90]{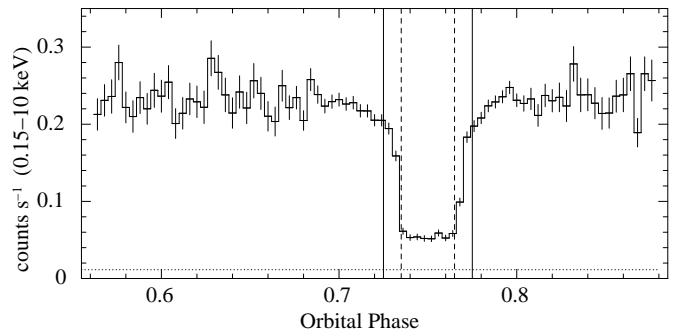}
\caption{\footnotesize {Light curve of \rx\ in the 0.15-10 keV energy range folded at the orbital period in bins of 535 s.
The light curve has been obtained by summing the pn and MOS data of the six observations of 2011 and of the 2008 observation. The horizontal
line indicates the background level. The vertical lines delimit the phase intervals used for the spectral analysis.}
 \label{fig_lc_orb}}
 \end{figure}

\section{Data analysis and results}

The 0.15--10 keV  light curve obtained by folding the 2008 and  2011 EPIC data (pn plus MOS) at the  $P=1.547667$ d orbital period
is shown in Fig.\ref{fig_lc_orb}.
The horizontal line shows the level of the background and the  vertical lines delimit the time intervals used for the spectral analysis described below.

\subsection{Emission during the eclipse}

\begin{figure}
 \includegraphics[width=65mm,angle=-90]{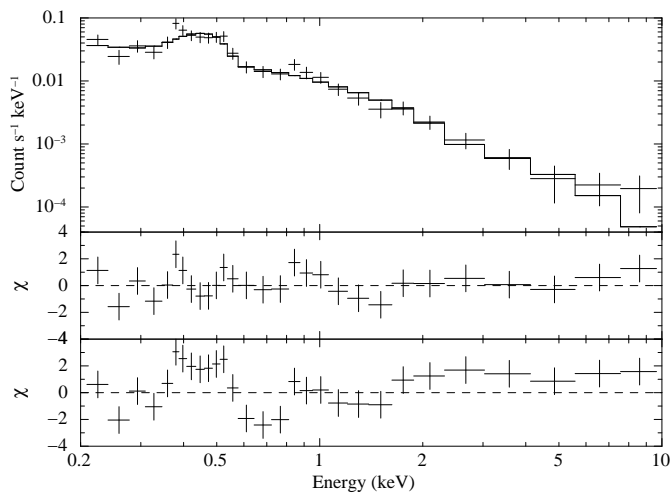}
\caption{\footnotesize {X-ray spectrum during the eclipse (pn  data).
Top panel: data and best-fit model consisting of
an absorbed power law and two Gaussian emission lines. Middle  panel:  residuals of the best-fit model.
Bottom panel: residuals obtained using only an absorbed power law. }
 \label{fig_spettro_ecl}}
 \end{figure}

We start  by examining the spectrum of the X-rays emitted during the eclipse,
which can now be studied in  detail thanks to the seven-fold
increase in exposure time compared to the previous observations.
The  spectrum of the single eclipse observed in May 2008   ($\sim$4 ks exposure) was well fit by either a thermal bremsstrahlung
($kT_{\rm Br}$= 0.55 keV) or a power law (photon index = 2.8), but these simple models are rejected by the inclusion of the new data.
In fact no simple single-component model can fit the total EPIC  spectrum (2008 + 2011, 0.2--10 keV) extracted from the 0.735--0.765 phase
interval  (vertical dashed lines in Fig.\ref{fig_lc_orb}).
In all cases  the reduced $\chi^2$ values are unacceptable
(2.8, 7.1 and 4.3 for  absorbed power law, blackbody and bremsstrahlung, respectively) and the  residuals  from the best-fit model
show the presence of a significant excess at 0.4--0.5 keV, probably due to a blend of unresolved emission lines
(lower panel of Fig.\ref{fig_spettro_ecl}).
A good fit is obtained with a power-law plus two narrow lines\footnote{We fixed the width to $\sigma=0$ keV,
which corresponds to the assumption of an intrinsic width smaller than the instrumental resolution.} with energies fixed at $E_{1}$=0.43 keV
and $E_{2}$=0.50 keV,
corresponding to  emission  from  N\textsc{vi} and N\textsc{vii} ions.
The  best-fit  spectrum of the pn   data  is shown in Fig.\ref{fig_spettro_ecl}.
Fully consistent results were obtained with the MOS spectra. The best-fit parameters obtained by the joint pn and MOS analysis
are given in Table \ref{tab_spettro_ecl}.

After background subtraction, the  RGS  spectra corresponding to the eclipse phases
contain only about 100  counts in total,  consistent with what expected from the best-fit EPIC model.
An excess of counts is present at the energy of the most intense line. Fitting this excess with
a narrow line with energy fixed at 0.431 keV, yields a line flux of ($1.1\pm0.7)\times10^{-5}$ ph cm$^{-2}$ s$^{-1}$.
No excess is seen at 0.5 keV, but this is not surprising considering the small photon counting statistics
of the RGS spectra and the low  flux estimated with EPIC for this line.

To explore the possibility that the X-rays during the eclipse originate in \hd\ we
tried a fit with a combination of plasma emission models with different temperatures.
The spectra of a large sample of O-type stars detected with \xmm\ were well described
by the sum of up to three thermal components\footnote{Model \texttt{MEKAL} in \texttt{XSPEC}.}
with temperatures from $\sim$0.1 to a few keV \citep{naz09}.
Using  this model with plasma Solar abundances we could not obtain a good fit to the pn spectrum.
This is not surprising considering the overabundances of N and He in this system \citep{kud78,ham10}.

\begin{figure}
 \includegraphics[width=65mm,angle=-90]{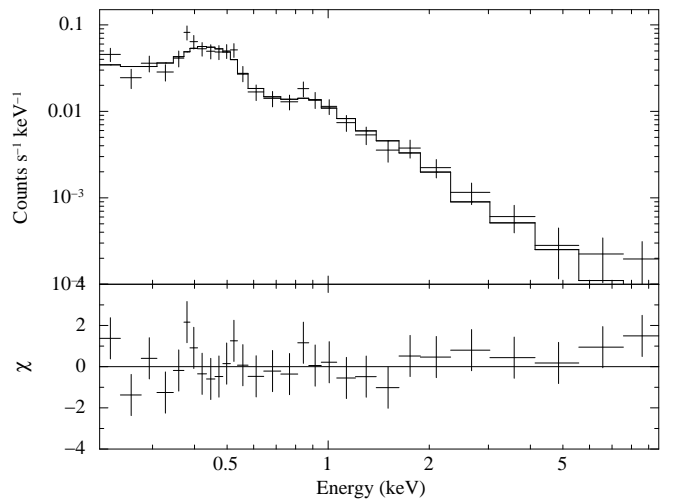}
\caption{\footnotesize {Spectrum during the eclipse (pn data) fit
with three thermal plasma emission  models. Bottom panel: residuals from the best-fit model. }
 \label{fig_3mekal}}
 \end{figure}

A good  fit (reduced $\chi^2=0.86$ for 23 d.o.f., Fig.\ref{fig_3mekal}) was instead obtained with the sum of three
thermal plasma models with the He and N abundances set to the values observed in \hd\  (mass fractions
of $X_{\mathrm{He}}=0.78$ and $X_{\mathrm{N}}=0.025$).
The derived temperatures are  $kT_1\sim0.14$,  $kT_2\sim0.7$, $kT_3=5$ keV\footnote{The hottest component is required to
account for the significant emission above $E>5$ keV. Its temperature is very
poorly constrained by the data,  so we fixed it at the plausible value $kT_3=5$ keV.}.
The 0.2--10 keV flux is $7\times10^{-14}$ erg cm$^{-2}$ s$^{-1}$.
This model is also consistent with the RGS data.

\subsection{Emission out of the eclipse}

We extracted from the six 2011 observations the spectra excluding the phase interval 0.725--0.775
(solid lines in Fig.\ref{fig_lc_orb}).
After checking that they  did not show statistically significant differences in flux or spectral parameters,
we summed them to obtain a total spectrum for each camera.
Assuming that the X-ray emission observed during the eclipse is a steady component present at all orbital phases,
as it was done also in \citet{mer11} for the 2008 data, we include in the  fit the
eclipse model with parameters fixed at the values of Table \ref{tab_spettro_ecl}.
In addition, two other components are required to fit the data: a blackbody
plus either a power law or a bremsstrahlung, yielding  the results summarized in Table  \ref{tab_spettro_unecl}.
Having  included a more accurate description of the eclipse spectrum compared to our previous works, we obtain now
slightly different  best-fit values for the non-eclipsed emission.
In particular,  the blackbody component derived in \citet{mer11} had  a higher temperature ($\sim$40 eV)
and a smaller emitting radius ($\sim$18 km), compared to the updated values reported here.

By summing the   2008 and 2011 RGS data corresponding to the uneclipsed
emission we obtained a spectrum with total exposure time of 129 ks. This spectrum suggests that the line at  0.431 keV
is present also out of the eclipse, since it shows  an excess with
flux of ($0.65\pm0.21)\times10^{-5}$ ph cm$^{-2}$ s$^{-1}$ (i.e. a significance $\sim$3$\sigma$).

\subsection{Eclipse duration and shape}

We fit the folded orbital light curve with a  piecewise function consisting of two constants (the eclipse and out-of-eclipse values) connected
by linear functions to model the eclipse ingress (between $\Phi^{\mathrm{i}}_1$ and $\Phi^{\mathrm{i}}_2$) and egress (between $\Phi^{\mathrm{e}}_1$ and $\Phi^{\mathrm{e}}_2$).
We assumed a symmetric profile, i.e. the same duration for the ingress and egress.
With a fit to the pn light curve in the 0.15--0.5 keV energy range, where the signal-to-noise ratio
is maximum, we obtained a  total eclipse duration ($\Phi^{\mathrm{e}}_1-\Phi^{\mathrm{i}}_2$) of $4311\pm52$ s
and a duration of  $525\pm49$ s  for the eclipse ingress/egress.
We  repeated the same analysis after dividing the  0.15--0.5 keV counts into two energy bands
with roughly the same number of counts without finding statistically significant evidence for any energy dependence of the eclipse parameters.

The gradual  ingress and egress phases can be explained by either assuming
that the X-rays originate from an extended region around the compact object or that they are gradually
absorbed in the densest regions  of the  wind close to the surface of \hd .
In the hypothesis of absorption in the  atmosphere/wind of \hd ,
we would expect to observe some energy dependence in the orbital light curves near the eclipse.
As mentioned above, the comparison of the eclipse profiles in
two energy bands was inconclusive, so we attempted to test this hypothesis
by analyzing the spectra  of the ingress and egress phases as follows.

We summed all the EPIC/pn spectra   extracted from  the phase intervals  0.725--0.735 and 0.765--0.775
and tried to fit the resulting spectrum by varying in two different ways the blackbody
plus power-law  model of the non-eclipsed emission.
This was done
a) by introducing  a multiplicative factor (to simulate the gradual occultation of an extended source)
and
b)  by keeping only the column density  as free parameter
(to check whether the flux decrease can be explained only by  absorption).
In both cases  all the other parameters were kept fixed at their best-fit values of the non-eclipsed emission
and the assumed constant eclipse spectrum (Table \ref{tab_spettro_ecl}) was included.
The resulting  values of reduced $\chi^2=0.99$ and $\chi^2=1.42$, respectively,  favor the first interpretation.

\subsection{Spin period evolution}

\begin{figure}
 \includegraphics[width=85mm,angle=-90]{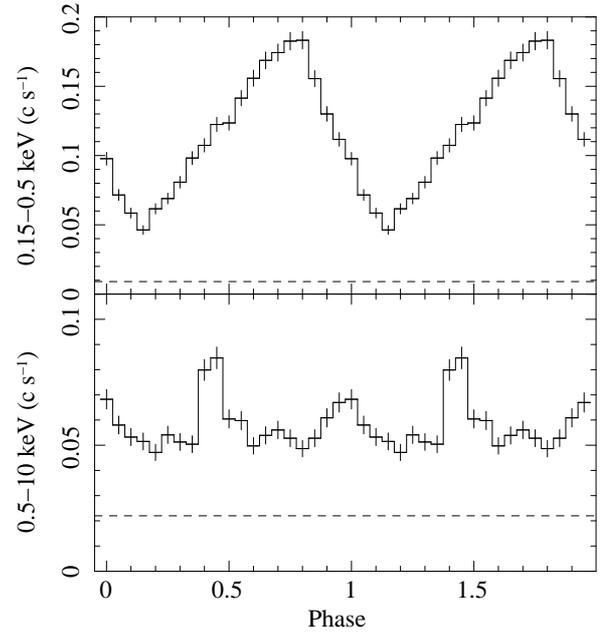}
\caption{\footnotesize {EPIC pn light curve of \rx\   folded at the spin period in the
soft (0.15--0.5 keV, top panel) and hard (0.5--10 keV, bottom panel) energy ranges. The dashed lines indicate the background level.}
 \label{fig_lc_spin2011}}
 \end{figure}

The   pulsations at the  spin period of 13.2 s
are clearly detected at consistent values   in each of the six  observations carried out in 2011.
To improve the accuracy of the period determination, we performed a joint timing analysis of the 2011 pn data
in the soft X-ray band ($<$0.5 keV) as follows.
We first phase-connected the two most closely spaced observations (n. 2 and 3) in order to obtain a preliminary solution.
This was then iteratively improved by the successive inclusion of the other pointings obtained at increasingly longer time intervals
(in the following order: 4, 5, 6, 1).
The  phase-connected ephemeris of the six observations gave a period  $P_{2011} = 13.18424736(16)$ s, which is
within the errors of the much less accurate value measured in May 2008  ($P_{2008} = 13.18425(4)$ s,  \citealt{mer11}).
A linear fit to these two values plus all the other period measurements (2002 May and September with \xmm\ and
November 1992 with $ROSAT$) gives the following limits on the period derivative: $-2.9\times10^{-13}<\pdot<3.6\times10^{-13}$ s s$^{-1}$  (90\% c.l.).
We obtained  a tighter  constraint on $\pdot$    by extending the  phase-connected timing analysis to the 2008 data and fitting the
pulse phases with a quadratic function:  the results are  $P = 13.1842474(2)$ s and
$|\pdot|<6\times10^{-15}$ s s$^{-1}$.
The 2011 light curves folded at the spin period are shown in Fig. \ref{fig_lc_spin2011} for the soft and hard energy ranges.
As in the previous observations, the light
curve   at low energy shows a strong nearly sinusoidal modulation (pulsed fraction $\sim$55\%), while two peaks are visible   above 0.5 keV.

\begin{table}[h]
\caption{EPIC spectral results for the eclipse emission.
\label{tab_spettro_ecl}}
\begin{tabular}{ll}
\hline\hline
 Parameter		& Value			\\
\hline
                        &     \\
$N_{\mathrm{H}}$  ($10^{20}$  cm$^{-2}$)		& $<$2.9 \\
Photon index		& $1.87_{-0.15}^{+0.22}$	\\
$F_{\rm PL}^{a}$  ($10^{-14}$  erg cm$^{-2}$ s$^{-1}$)	& $7.3\pm0.7$  \\
$E_{1}$ (keV) 		& 0.43   (fixed)\\
$I_{1}$  ($10^{-5}$  ph cm$^{-2}$ s$^{-1}$)  		& $1.1_{-0.4}^{+0.7}$ \\
$E_{2}$ (keV)   		& 0.5 (fixed)\\
$I_{2}$ ($10^{-5}$  ph cm$^{-2}$ s$^{-1}$ )		& $0.68_{-0.25}^{+0.29}$ \\
d.o.f.			& 35				\\
$\chi^{2}_{\nu}$	& 1.16				\\
\hline
\end{tabular}
\begin{small}
\\
$^{a}$ Flux of the power-law component in the 0.2--10 keV range.
\\
\end{small}
\end{table}

\begin{table}
\caption{EPIC spectral results for the uneclipsed emission.
\label{tab_spettro_unecl}
}
\begin{tabular}{lccc}
\hline\hline
                              & BB+PL         & BB+Brem \\
\hline
 & &  \\
$N_{\mathrm{H}}$  ($10^{20}$ cm$^{-2}$)				&  $1.6\pm0.7$   &    $0.5\pm0.5$        \\
$kT_{\rm BB}$ (eV)					& $30.6\pm1.5$	 &      $33.5\pm1.3$    \\
$R_{\rm BB}^{a}$ (km)					& $66_{-46}^{+67}$	&     $35_{-22}^{+30 }$     \\
Photon index		 				& $1.96_{-0.10}^{+0.11}$	&       -- \\
$F_{\rm PL}^{b}$ ($10^{-13}$ erg cm$^{-2}$ s$^{-1}$)	& $1.01_{-0.03}^{+0.05}$	&        --   \\
$kT_{\rm Br}$	 (keV)					& 	--              & $3.7_{-0.7}^{+0.9}$ \\
$F_{\rm Br}^{b}$ ($10^{-13}$ erg cm$^{-2}$ s$^{-1}$)	& 	-- 		&$0.78_{-0.05}^{+0.06}$\\
d.o.f.							&  262		&	262	\\
$\chi^{2}_{\nu}$					&  1.27		&  	1.42\\
\hline
\end{tabular}
\begin{small}
\\
$^{a}$ Blackbody emission radius for $d=650$ pc.
\\
$^{b}$  Flux in the range  0.2--10 keV.
\\
\end{small}
\end{table}

\section{Discussion and conclusions}

The presence of X-ray emission also during the eclipse, already noticed in  May 2008,  is
confirmed by the new observations of \hr , which show a spectrum characterized by strong N emission lines.
A possible explanation is that these  X-rays   originate in \hd\ itself, i.e.
they are unrelated to the presence of a compact object in this system.
This is particularly interesting since no O-type subdwarf has been detected in X-rays up to now
\footnote{with the   exception of BD\,+37$^\circ$\,442  \citep{lap12}.
However, the possible presence of pulsations at 19 s suggests that also BD\,+37$^\circ$\,442 has a compact companion.}.

Our current understanding of the X-ray emission from early type stars is based on OB main sequence, giant
and supergiant stars, which, compared to \hd ,  have very different luminosity, mass  and composition.
The X-ray emission of these stars originates in their strong, radiation-driven  winds, where  plasma is heated by shocks and
instabilities \citep{pal89}.
High resolution spectra of bright O type stars are characterized by the presence of numerous emission lines (see, e.g. \citealt{kah01}),
but data  with lower resolution and/or
statistical quality are usually well fit by a combination of plasma emission models with different temperatures.
The temperatures we derived are fully consistent with those
found in a large sample of O stars studied with \xmm\  \citep{naz09}.
The luminosity of $\sim$$3\times10^{30}$ erg s$^{-1}$ that  we measured during the eclipse corresponds to an X-ray-to-bolometric flux ratio
of    $\sim$$10^{-7}$, as typical  O-type stars \citep{pal89}.
Thus we  conclude that the X-ray luminosity and spectrum  observed during the eclipse are consistent
with those of early type stars
and support the interpretation of \hd\ as the first hot subdwarf detected at X-ray energies.

In this respect it is interesting to note that \hd\ is one of the few sdO stars for which evidence of wind
mass loss has been reported \citep{ham81}.
This suggests that the processes responsible for X-ray emission in the stellar winds
of   massive O stars are also at work in the much weaker winds of hot subdwarfs
and makes \hd\ a particularly interesting target
to explore the properties of radiatively driven winds through X-ray observations over a wide  range of stellar parameters.

Thanks to the better coverage of the eclipse orbital phases, the new observations allowed us  to study the eclipse shape.
We derived a  duration of 4311$\pm$52 s for the total eclipse,  consistent with the rough value of 4300 s
that was estimated from the 2008 observation and used to constrain the system inclination in \citet{mer09}.
We could also establish that the eclipse ingress and egress occur gradually. This can be explained if
part of the X-ray emission originates in an extended region surrounding the compact object.
It is easy to show with a simple geometrical model, based on the known
orbital parameters and radius of \hd\ (1.45 $R_\odot$, \citealt{kud78}), that the observed duration of $\sim$500 s of the eclipse
ingress and egress
implies dimensions of the order of a few 10$^4$ km,  much larger than the radius of a massive white dwarf ($\sim$3000 km).
The presence of strong pulsations, which must necessarily originate close to the compact  object, sets a limit on the
flux coming from the extended region.  Considering that $\sim$55\% of the counts below 0.5 keV are pulsed,
this limit is of about $2\times10^{-13}$ erg cm$^{-2}$ s$^{-1}$  (0.2--0.5 keV).
A plausible scenario is that part of the high-energy radiation emitted by the compact object is
scattered in the wind of \hd , as observed in more luminous X-ray binaries powered by wind accretion,
such as, e.g., Vela X-1  or  Cen X-3  \citep{sak99,ebi96}. This interesting possibility
deserves to be explored with more sensitive X-ray observations with good spectral resolution (Sartore et al. in preparation).

The new observations confirm the stability in the long term properties of the X-ray emission from this binary system:
the  luminosity and pulse profile do not show significant variations compared to previous observations,
but thanks to a better modeling of the eclipse spectrum we could obtain more accurate values
for the spectral parameters of the uneclipsed emission.
The lower temperature of the blackbody-like component, which dominates
the low-energy pulsed emission, implies an emitting area difficult to reconcile with origin on
the surface of a neutron star.

Thanks to a properly planned spacing of the observations, we could obtain for the first time
a phase-connected timing solution for this system.
This provides stronger constraints on the secular evolution of the 13 s spin period:
the new limit  on the period derivative, $|\pdot|<6\times10^{-15}$ s s$^{-1}$ (90\% c.l.), is more than two orders of magnitude smaller
than the previous value (--5$\times$10$^{-13}$ s s$^{-1}$ $<\pdot<$ 9$\times$10$^{-13}$ s s$^{-1}$, \citet{mer11}).
Considering the small torque expected for the low accretion rate onto this pulsar, this limit is still
insufficient  to definitely rule out  a neutron star, a possibility not dismissed yet, although we believe that
the observed X-ray luminosity and spectrum are best explained by a white dwarf companion.
In this respect it is essential to   extend the
phase-connected solution over a longer time span by means of new X-ray measurements of the spin period.

\begin{acknowledgements}
This work is based on observations obtained with \xmm , an ESA science mission
with instruments and contributions directly funded by ESA Member States and NASA.
We acknowledge financial contributions by the Italian Space Agency through ASI/INAF
agreements I/009/10/0 and I/032/10/0  and by INAF through PRIN INAF 2010.

\end{acknowledgements}

\bibliographystyle{aa}

\begin{thebibliography}{18}
\expandafter\ifx\csname natexlab\endcsname\relax\def\natexlab#1{#1}\fi

\bibitem[{{den Herder} {et~al.}(2001){den Herder}, {Brinkman}, {Kahn},
  {Branduardi-Raymont}, {Thomsen}, {Aarts}, {Audard}, {Bixler}, {den Boggende},
  {Cottam}, {Decker}, {Dubbeldam}, {Erd}, {Goulooze}, {G{\"u}del}, {Guttridge},
  {Hailey}, {Janabi}, {Kaastra}, {de Korte}, {van Leeuwen}, {Mauche},
  {McCalden}, {Mewe}, {Naber}, {Paerels}, {Peterson}, {Rasmussen}, {Rees},
  {Sakelliou}, {Sako}, {Spodek}, {Stern}, {Tamura}, {Tandy}, {de Vries},
  {Welch}, \& {Zehnder}}]{denHerder+01long}
{den Herder}, J.~W., {Brinkman}, A.~C., {Kahn}, S.~M., {et~al.} 2001, \aap,
  365, L7

\bibitem[{{Ebisawa} {et~al.}(1996){Ebisawa}, {Day}, {Kallman}, {Nagase},
  {Kotani}, {Kawashima}, {Kitamoto}, \& {Woo}}]{ebi96}
{Ebisawa}, K., {Day}, C.~S.~R., {Kallman}, T.~R., {et~al.} 1996, \pasj, 48, 425

\bibitem[{{Hamann}(2010)}]{ham10}
{Hamann}, W. 2010, \apss, 119

\bibitem[{{Hamann} {et~al.}(1981){Hamann}, {Gruschinske}, {Kudritzki}, \&
  {Simon}}]{ham81}
{Hamann}, W., {Gruschinske}, J., {Kudritzki}, R.~P., \& {Simon}, K.~P. 1981,
  \aap, 104, 249

\bibitem[{{Israel} {et~al.}(1997){Israel}, {Stella}, {Angelini}, {White},
  {Kallman}, {Giommi}, \& {Treves}}]{isr97}
{Israel}, G.~L., {Stella}, L., {Angelini}, L., {et~al.} 1997, \apjl, 474, L53

\bibitem[{{Kahn} {et~al.}(2001){Kahn}, {Leutenegger}, {Cottam}, {Rauw},
  {Vreux}, {den Boggende}, {Mewe}, \& {G{\"u}del}}]{kah01}
{Kahn}, S.~M., {Leutenegger}, M.~A., {Cottam}, J., {et~al.} 2001, \aap, 365,
  L312

\bibitem[{{Kudritzki} \& {Simon}(1978)}]{kud78}
{Kudritzki}, R.~P. \& {Simon}, K.~P. 1978, \aap, 70, 653

\bibitem[{{La Palombara} {et~al.}(2012){La Palombara}, {Mereghetti}, {Tiengo},
  \& {Esposito}}]{lap12}
{La Palombara}, N., {Mereghetti}, S., {Tiengo}, A., \& {Esposito}, P. 2012,
  \apjl, 750, L34

\bibitem[{{Mereghetti} {et~al.}(2011){Mereghetti}, {La Palombara}, {Tiengo},
  {Pizzolato}, {Esposito}, {Woudt}, {Israel}, \& {Stella}}]{mer11}
{Mereghetti}, S., {La Palombara}, N., {Tiengo}, A., {et~al.} 2011, \apj, 737,
  51

\bibitem[{{Mereghetti} {et~al.}(2009){Mereghetti}, {Tiengo}, {Esposito}, {La
  Palombara}, {Israel}, \& {Stella}}]{mer09}
{Mereghetti}, S., {Tiengo}, A., {Esposito}, P., {et~al.} 2009, Science, 325,
  1222

\bibitem[{{Naz{\'e}}(2009)}]{naz09}
{Naz{\'e}}, Y. 2009, \aap, 506, 1055

\bibitem[{{Pallavicini}(1989)}]{pal89}
{Pallavicini}, R. 1989, \aapr, 1, 177

\bibitem[{{Perryman} \& {ESA}(1997)}]{per97}
{Perryman}, M.~A.~C. \& {ESA}, eds. 1997, ESA Special Publication, Vol. 1200,
  {The HIPPARCOS and TYCHO catalogues. Astrometric and photometric star
  catalogues derived from the ESA HIPPARCOS Space Astrometry Mission}

\bibitem[{{Sako} {et~al.}(1999){Sako}, {Liedahl}, {Kahn}, \& {Paerels}}]{sak99}
{Sako}, M., {Liedahl}, D.~A., {Kahn}, S.~M., \& {Paerels}, F. 1999, \apj, 525,
  921

\bibitem[{{Str{\"u}der} {et~al.}(2001){Str{\"u}der}, {Briel}, \&
  {Dennerl}}]{str01}
{Str{\"u}der}, L., {Briel}, U., \& {Dennerl}, K.~e. 2001, \aap, 365, L18

\bibitem[{{Thackeray}(1970)}]{tha70}
{Thackeray}, A.~D. 1970, \mnras, 150, 215

\bibitem[{{Turner} {et~al.}(2001){Turner}, {Abbey}, {Arnaud}, \& {et
  al.}}]{tur01}
{Turner}, M.~J.~L., {Abbey}, A., {Arnaud}, \& {et al.} 2001, \aap, 365, L27

\bibitem[{{Wang} \& {Han}(2010)}]{wan10}
{Wang}, B. \& {Han}, Z. 2010, Research in Astronomy and Astrophysics, 10, 681

\end{thebibliography}

\end{document}